# GCUBE INDEXING

M.Laxmaiah[1] and A.Govardhan[2]

[1]Department of Computer Science and Engineering,
Tirumala Engineering College, Bogaram (V), Keesara(M), Hyderabad,
AP-, India-501301, `laxmanmettu.cse@gmail.com`
[2]School of Information Technology,
Jawaharlal Nehru Technological University, Hyderabad,
AP, India-500085, `govardhan_cse@yahoo.co.in`

## ABSTRACT

*Spatial Online Analytical Processing System involves the non-categorical attribute information also whereas standard online analytical processing system deals with only categorical attributes. Providing spatial information to the data warehouse (DW); two major challenges faced are; 1.Defining and Aggregation of Spatial or Continues values and 2.Representation, indexing, updating and efficient query processing. In this paper, we present GCUBE (Geographical Cube) storage and indexing procedure to aggregate the spatial information/Continuous values. We employed the proposed approach storing and indexing using synthetic and real data sets and evaluated its build, update and Query time. It is observed that the proposed procedure offers significant performance advantage.*

## 1. INTRODUCTION

Indexing of multi-dimensional point is a well studied topic. The R-Tree [1] is a spatial indexing structure that allows the indexing and efficient query of points, polygon shapes and it is considered to be the most commonly recommended and accepted indexing. The basic difference between traditional OLAP System and SOLAP System is that features in OLAP are categorical. Features can exploit in several ways, such as categorical data stored in *Star Schema as shown in* (Figure 1).

Categorical dimensions can be normalized in each dimension leading to the importance of fixed cardinality. The basic purpose of defining dimension hierarchy is to support drill down and roll up operations. It is not compulsory that every captured data is categorical or structured, unstructured data is also captured often. It is not compulsory that every captured data is categorical or structured, un structured data is also captured often Dealing with non categorical dimensions is real challenge for researcher and developers. The key issues arise when dealing with non categorical dimensions are: 1.Definition and aggregation on spatial/non-spatial / continuous data.2. Its representation, indexing, updating and efficient replies to the queries involving both kinds of the data [2, 3].Enumerating continuous dimensions as if they were values of categorical Space is followed for updating and indexing. This describes the original continuous dimension in data dependent manner and once continuous dimension have been transformed in this way, recompilation of the complete view becomes mandatory to view updates with respect to the original Space Research and business interest in OLAP with mixed dimensions has proposed various representations and indexing procedures for various things such as sequential [16, 4], parallel [22] and peer to peer [7]. In this paper, we propose GCUBE for representation and indexing of mixed dimensions in a Relational OLAP setting. Our proposed approach is intended for the indexing of views in ROLAP setting as the resulting data structure only represents those sections of the multidimensional view that data records associated with them.

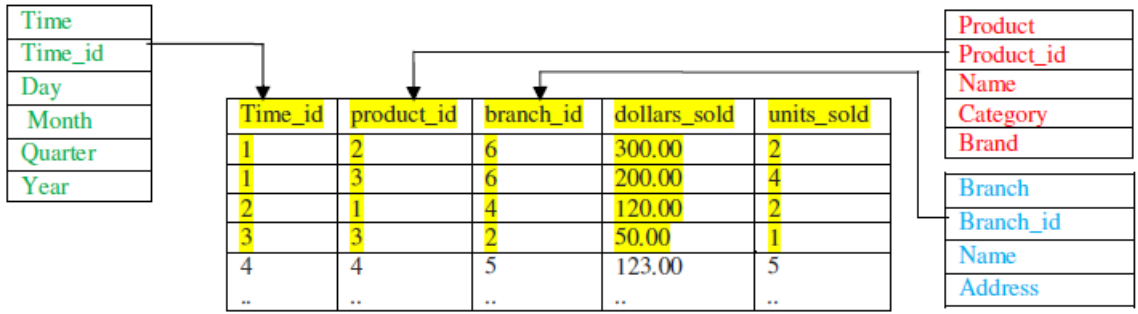

Figure 1: Star Schema of a typical data warehouse with facts table, three feature dimensions and two measures. The feature dimensions have been normalized within the facts table in that they are represented as dimension tables.

The proposed approach is employed and tested using synthetic and real time data. Emperical results imply that Hilbert curve based indexing improves the performance over 20-25%.It also over performed the standard pre-descretization approach. It is also observed that the indexing using Hilbert curve consumes less time for updates i.e. between a factors of 23 from an update of size 2% which is significant improvement. The remaining paper is organized as follows; section 2 presents the brief over view of the related work. Section 3 presents our basic approach followed by description of our approach for dynamically adapting the representation of the Hilbert curve in section 4.The orientation of index and its representation in memory is described in section 5,Section 6 presents the implementation issues and section 7 concludes the paper.

## 2. RELATED WORK

Over the many number of years, the problem of indexing multidimensional data has been studied and the most commonly employed approach for effective indexing is R-Tree and its variants like R+-Tree, R*-Tree, SS-Tree and SR-Tree. The R-tree [1] is a spatial indexing structure that allows the indexing and efficient querying of points and polygonal shapes. Real time range queries are supported by R-Tree and it is inspired by B-Tree. R-Tree also works well with external memory data. Good number of extensions to the R-Tree has been proposed to deal with various forms of spatial and spatio-temporal OLAP [6,8, 9, 10].Most approaches based on R-trees, however, focus only on continuous space dimensions and do not specifically address the properties of categorical dimensions. Categorical dimensions are typically transformed into an equivalent continuous Space. This transformation of categorical dimension to continuous space carries the problem with it. Problems could be calculation, methodology, complexity which intern lead to the difficulty of updates. The efficient aggregation in OLAP queries over categorical dimensions often crucially relies on these dimensions being integer-valued, and the use of space-filling curves as a locality-preserving mapping from higher-dimensional space into one dimension.

This approach organizes multi-dimensional categorical OLAP views by using the Hilbert space-filling curve to generate a linear ordering of the records in multi-dimensional space and then indexing this linear ordering with a data structure similar to a B-tree. This method exploits that the Hilbert curve strongly preserves spatial locality [12].Previously, Hilbert curves were used by Kamel and Faloutsos [13] as a mechanism to enhance the performance of R-trees indexing with multi dimensional data. Space-filling curves have been used to support the indexing of purely categorical OLAP data .Hilbert curve is employed by them to obtain the ordering the records with in each node of an R-Tree. This allows the exercise of new strategies for the distribution of records when nodes split and merge during insertion and deletion of records without compromising the Query performance of R-Tree. By using Hilbert curve tree structure Lewder and King proposed a multidimensional index [32].Each level of this tree corresponds to a level of resolution, or order of the Hilbert curve and partitions the covered subspace into quadrants. Records are stored at the leaf level of the tree in the subspace quadrants that correspond to the Hilbert values derived from their original coordinates and the entire space is covered at root. No work has been reported previously which uses Hilbert curve for indexing multidimensional and continuous data. All proposed techniques operate on multi-dimensional grids whose resolutions are known in advance, and records are mapped into cells of these grids. Hence, the cardinality of each dimension is necessarily being fixed and known in advance. However, it is not practical in a spatial OLAP environment where multiple continuous dimensions may exist and their attribute values are not fixed and may change over any number of updates. A more application-oriented approach to the integration of spatial and categorical data based on building composite systems that integrate existing OLAP and Geographical Information Systems has been pursued by both academic [11,12,13,14,15] and industrial [16,17,18, 19, 20] research groups.

## 3. BASIC APPROACH

The objective of this paper is to address the issue of the representation and indexing of mixed data is a ROLAP setting. This approach fails leading when massive data is available to work with. Sequential [4], parallel [23] and peer to peer [7] are sure of the employed representation and indexing techniques. Then the indexing is analyzed based on one space filling curves can be extended to mixed dimension setting. It has three basic steps;

1. Mapping of Multi-Dimensional Data into Linear Ordering using Hilbert Curve [21].
2. Use the Linear Order to distribute the data over the available storage.
3. Construct an index structure on top of the Ordered Data for efficient Query Processing [23, 1].

R-Tree is being generated; however one of space filling curves to fold the multi-dimensional space into a one dimensional space well suited for even distribution of records over multiple disks processors or peers. It facilitates the batch updates into the original view with a single linear scan.

- Sequential block access reduces the amount of disk I/O and seeks.
- In multidimensional setup no dimension is given favored.
- Extension to indexing provides additional support for OLAP.

International Journal of Data Mining & Knowledge Management Process (IJDKP) Vol.3, No.4, July 2013

The proposed approach uses Hilbert Curve as the space filling curve that indexes our method. The Hilbert Curve is defined as a *d*-Dimensional grid with a side length of $2^k$, where *d*, the number of dimensions of the view and *k* is the resolution or order of Curve. The grid covers the entire space of the view and the resolution *k* amount of closure in such a way that each grid cell contains at most one record as presented in figure2 below.

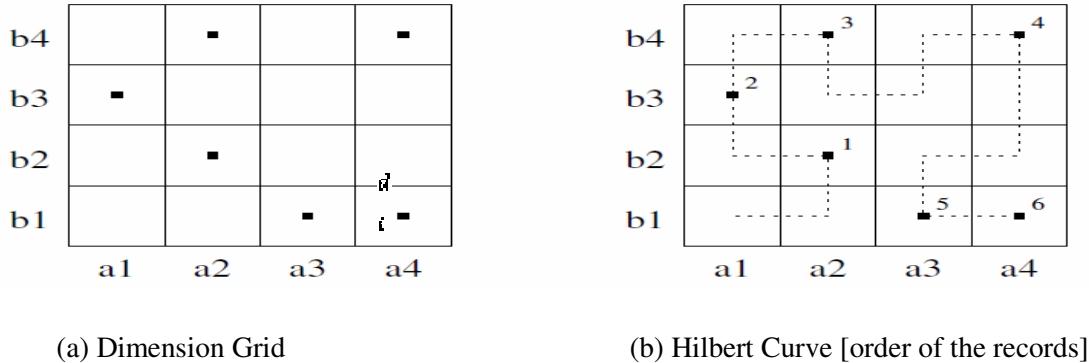

(a) Dimension Grid  (b) Hilbert Curve [order of the records]

Figure 2: Hilbert curve

Given a view with *d* categorical dimensions and each dimension *Di* containing |*Di*| distinct values (dimension cardinality) then a grid with a resolution of $k= [\log_2 \max |Di|]$ ensures each record is located in a distinct grid cell. The Hilbert curve on the grid passes through each record in the view exactly once and there by generates the Hilbert order of the records (Figure 2.b).

Hilbert curve provides the facility of mapping d-dimensional Space to 1-dimesional Space without losing the information of higher dimension [4].This new approach is flexible with respect to the resolution of the grid as it attempts to optimally utilize the grid space for continuous dimensions. It also reduces the grid cells compare to pre-descretization approach and new records are allowed with previously unknown attribute values without recomputing the entire view. Pair wise exclusion is necessary for predetermination of Grid resolution and duration time between records plays most crucial role when dealing with large records. Based on the recursive definition and the self-similarity of the Hilbert Curve, the Hilbert Ordering of records can be efficiently computed in continuous space by using an adaptive resolution for the underlying descretization grid. It is observed that the relative order of records is preserved by Hilbert ordering when the resolution of the Hilbert curve is increased (Figure 3)

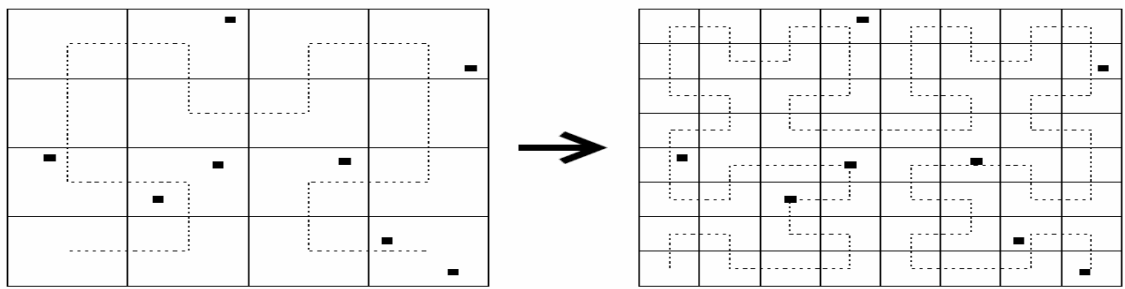

Figure 3: Increasing the resolution of the Hilbert Curve preserves the order of records.

Increase in the resolution of the Hilbert Curve is directly proportional to the resolution of the multi dimensional grid on which the curve id defined. This property is used to generate the distinct records map to distinct grid cells. Because of the Grid's dynamic nature the Space is not skewed and relative difference between continuous values of records is preserved. This property can be used to dynamically adjust the resolution of the grid on to which the continuous space attributes are mapped during the generation of Hilbert ordering of the records, in order to generate that distinct records map to distinct grid cells and thus become comparable in Hilbert order. Furthermore, this approach achieves a better space utilization than a pre-descretization approach, became a grid cell is split into smaller cells only if there is in fact a conflict between two records in the cell. Because of the Grid's dynamic nature the space is not skewed and relative difference between continuous values of records preserved.

Algorithm 1 outlines a method to produce a Hilbert ordering of records that are defined in with both continuous and discrete space. It utilizes a standard comparison based sorting algorithm such as Quick sort and Merge sort comparison between (algorithm 2) that determines the relative order of two records on a Hilbert curve at particular resolution. The relative order is estimated by finding the minimum resolution of the Hilbert curve such that both records have the different rank with respect to the curve.

---

Algorithm 1  Algorithm to sort records in Hilbert order using dynamic resolution adaptation

---

Procedure : Hilbert-sort

**Input**: set R of records in no particular order, set of categorical dimensions $D_i$, with $|D_i|$ being the number of distinct values in dimension $D_i$

**Output**: set R of records in Hilbert order

> ▷ Compute initial resolution

1. k ← $\lceil \log_2 \max\{|D_i|: \forall D_i \in D\} \rceil$

   ▷ Call sorting algorithm with Hilbert-**compare** as

2. **Sort**(R, Hilbert-**compare**$_k$ )

---

Algorithm 2  Algorithm to determine the order of two records with respect to the Hilbert curve by dynamically adapting the resolution of Hilbert curve if necessary

---

Procedure: Hilbert-compare

**Input**: pair of records (*r1, r2*), initial resolution k

**Output**: -1 if the $\text{rank}_k(r1) < \text{rank}_k(r2)$, 0 if (*r1=r2*) or 1 if $\text{rank}_k(r1) > (r2)$

1. if *r1=r2* then
2. return 0
3. else
4. while $\text{rank}_k(r1) + \text{rank}_k(r2)$ do
5. $k \leftarrow k+1$    determine resolution suitable for comparison

6. end while
7. if rank$_k$(*r1*) <  rank$_k$(*r2*) then
8. return  -1
9. else
10. return 1
11. end if
12. end if

Initial resolution is computed and formed during comparisons and the resolution is increased accordingly. Therefore the resolution of the grid is either maintained or concerned from each comparison. After the sorting process, all records are in Hilbert order and the resolution determined by the last comparison during the sorting process is the minimal resolution for a grid such that every record of the dataset maps to a distinct cell in the grid.

## 3.1 Efficient Indexing

Once the Hilbert ordering of the records has been determined, the records are sequentially written to disk in a block-wise manner. Each block on disk stores a constant number B of records that are consecutive in Hilbert order. Over the ordered records an indexing structure is formulated which provides features comparable to those of a combination of B-Tree of an R-Tree [23] Figure 4 illustrate the construction of such a tree structure with specific example

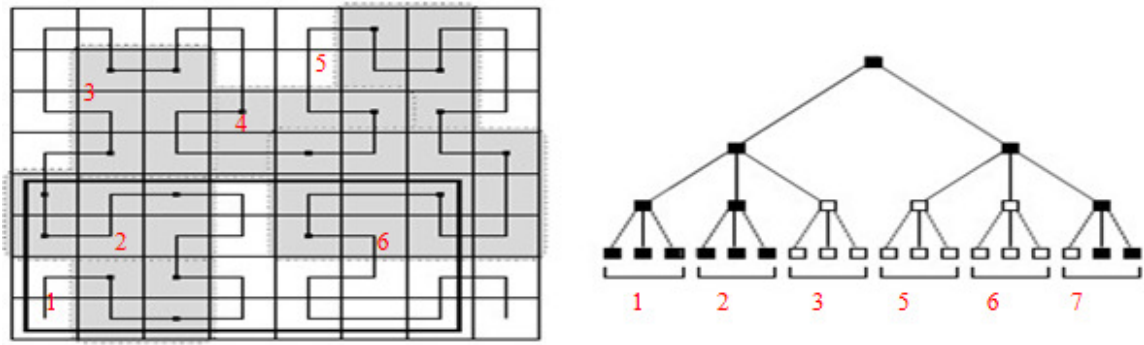

Figure 4: Records in Hilbert order are stored in consecutive blocks on disk and from multi- dimensional regions in the original space. The evaluation of a range query in breadth- first fashion only traverses part of the tree and results in combination of sequential read and forward-seek operations at the leaf level.

While each intermediate node of the tree is very similar to node in conventional B-Tree. It is also annotated with minimum bounding box of the records in its sub tree, similar to nodes in R-Tree. Due to the self similarity of the Hilbert curve and its properties to  not favor   any specific dimensions, the  bounding boxes of the records in each level  of indexing structure  are usually flat and mutually overlap only very little and helps in improving performance of the query.

Tree is traversed in BFM to respond a Query, by limiting the disk access to a combination of sequential records, forward seek operations and reducing the amount of random access [11].At time the discussed indexing has to answer range aggregate Queries. This is supported with direct reference sequence of leaves in its sub tree. To re-traverse the acquired records, the Tree is traversed by starting from root and determining each child nodes of non –leaf node, if it's bounding box intersects with query region. If so, all in some records in its sub tree may constitute to the final query result, so the traversal continuous in the sub tree of this node. The re-traversal of all records contained in a range to compute a single aggregate value is, however computationally expensive and is often overkill because the records only constitute to the aggregate value and are discarded after this computation.

To help aggregate OLAP queries efficiently, the aggregate information to be stored at each node is pre computed, while iterating over the children of the node during the bottom up construction of index tree. Note that this pre-computation of aggregate information works well for distributive (e.g. COUNT, SUM) and algebraic (e.g. AVERAGE) aggregation functions. Holistic aggregation functions (e.g. MEDIAN) on other hand require the retrieval of the actual records enclosed in the query region in order to compute the aggregate value. This is supported with direct reference sequence of lines in its sub tree. When evaluating a query the records in a node's sub tree can be reported immediately once the node has been reached, without traversing remaining sub tree below this node (Refer Figure5)

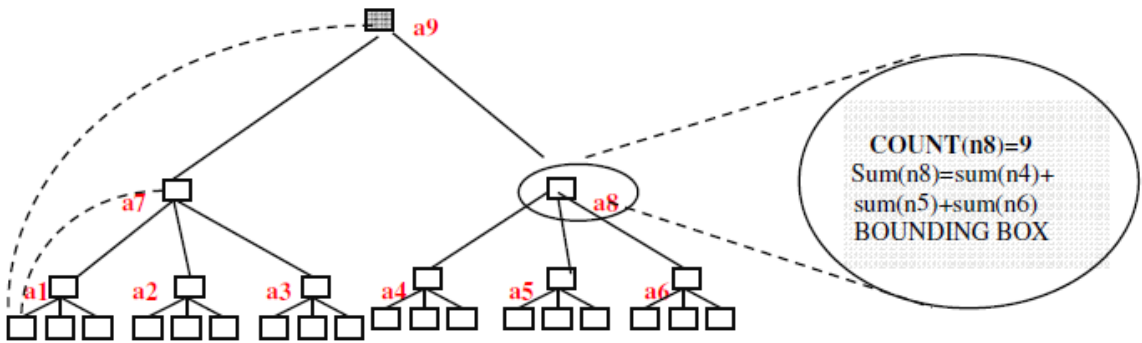

Figure 5: Annotation of non-leaf nodes with aggregate information and references to records at the leaf level.

## 3.2 Updating View

Once the update records are in the Hilbert order, they are merged with original dataset in a single scan over both data sets. Algorithm 3 shows in detail how both sequences are merged.

Algorithm 3 Merge algorithm to incorporate update dataset into target dataset.

Procedure: Hilbert-merge

**Input**: stream U of update records in Hilbert order, stream T of existing target records in Hilbert order, resolution k of the Hilbert curve used to order the existing records

**Outpu**t: merged stream O of records in Hilbert order

1. let u be the first element in U or empty if no such element exists
2. let t is be first element in T or empty if no such element exists
3. While u and t are not empty do

4. if Hilbert- compare (u, t ,k)=0 then
5. Write u to O
6. Let t be the next element in the T or empty if no such
    element exists
7. Let u be the next element in the U or empty if no such
    element exists
8.  else if Hilbert-compare (u, t, k) <0 then
9.  Write u to O
10. let u be the next element in U or empty if no such element   exists
11. else
12. Write t to O
13. let t be the next element in T or empty if no such element exists
14. end if
15. end while
16. if u is not empty then
17. Write the remainder of U to O
18. else if t is not empty then
19. Write   the remainder of T to O
20. end if

It iterates through the record in both datasets in Hilbert order and repeatedly compares the two current records from both data sets. If both the records shares the same attribute values, the records from the update dataset is considered an update of the existing record and therefore moved into the output data P, while the original record is discarded. If the attribute values of both record differ in at least one dimension, the Hilbert ranks0 of both records at the initial resolution are determined and compared. If one record has lower Hilbert rank than the other, it is moved to the output dataset and new current record is fetched from the respective input dataset. Since the locality of the records mapped into the Hilbert space is preserved when increasing the resolution of the Hilbert curve, records that have been merged already are not affected by the increase of resolution (Figure6). The merging process continuous until all records from both data sets has been processed. The output data set is a set of records in Hilbert order and will contain all records from the original data set that have not been updated.

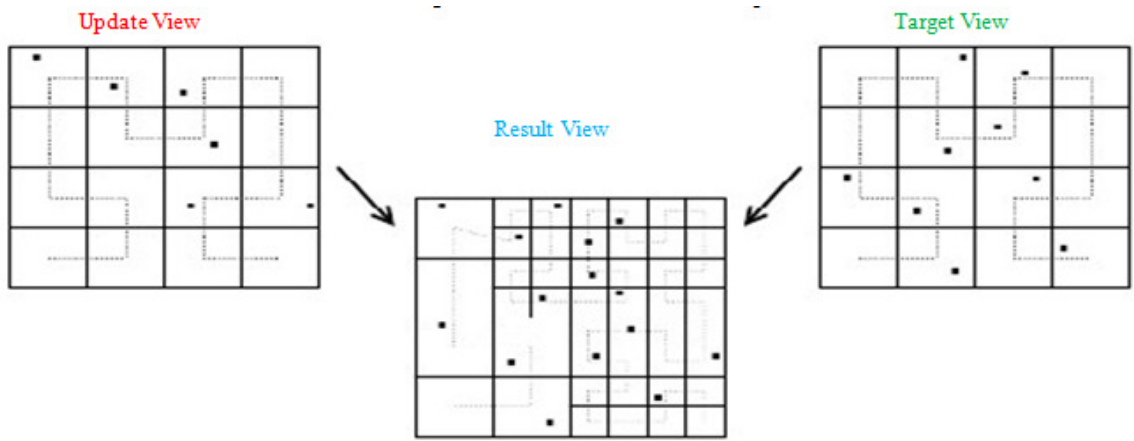

Figure 6: Merging the target view with the update view may result in a dynamic adaptation of the Hilbert curve resolution

## 4. IMPLEMENTATION AND EMPIRICAL ANALYSIS

The Proposed approach is well suited for determining the Hilbert order from records with dimensions defined in continuous and discrete space. It also works well on query evaluation and reduces update time. Modified Hilbert Space brings during Moore's Hilbert mapping library is implemented for mapping from n-dimensional discrete space to one dimensional Hilbert space [27].

### 4.1 Performance Evaluation

The experiments conducted to evaluate the cost of sorting records into Hilbert order, building the index, evaluating queries, and performing view updates. Experiments are carried out using synthetic and real time data. Generation of syntactic data is carried out according to the information distribution which implements the effects of various parameters on performance. The categorical and outcome dimensions of the synthetic datasets are generated with consideration of 64 and 1000, respectively. The real world datasets are drawn from the HYDR 01K dataset published by the US. Standard preprocessing is done before considering the available data for training such as dimensionality and size reduction using information random Sampling. Test data available is composed of 6- dimensions of which 2 are categorical dimensions with 51 digital categorical values, and form continuous dimensions with 56,1000,802 and 212 distinct continuous values. Measured considered for evaluation of various algorithms the impact of the size of the datasets on the running times of the various algorithms. Unless stated otherwise all experiments are performed with the most beneficial optimization enabled. In particular, each record was annotated with its last computed Hilbert rank.

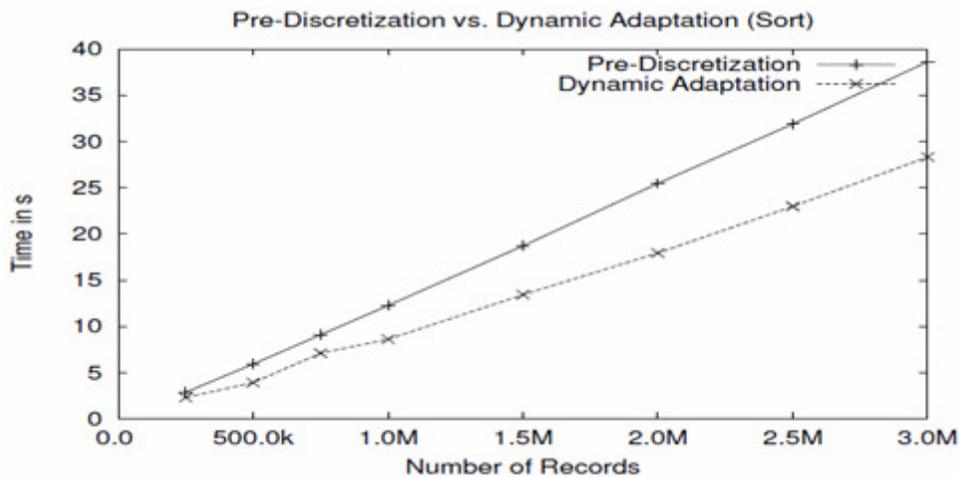

Figure 7: Comparing dynamic adaptive of the Hilbert curve versus traditional method of pre-descretization.

Empirical results imply that dynamic adaptation of the Hilbert curve performed pre descretization approach. It is observed that for similar set of records pre-descretization took approximately 38 seconds to pre-discretize and sort the data where the proposed approach took only 28 seconds to serve our purpose. This yield 26% more speed processing which can also be observed from the results for most other data sets tested. Figure7 shows the average range query time on HYDRO1K dataset over 1000 experiments using Geo CUBE index and R-TREE *index from spatial index library* [31].

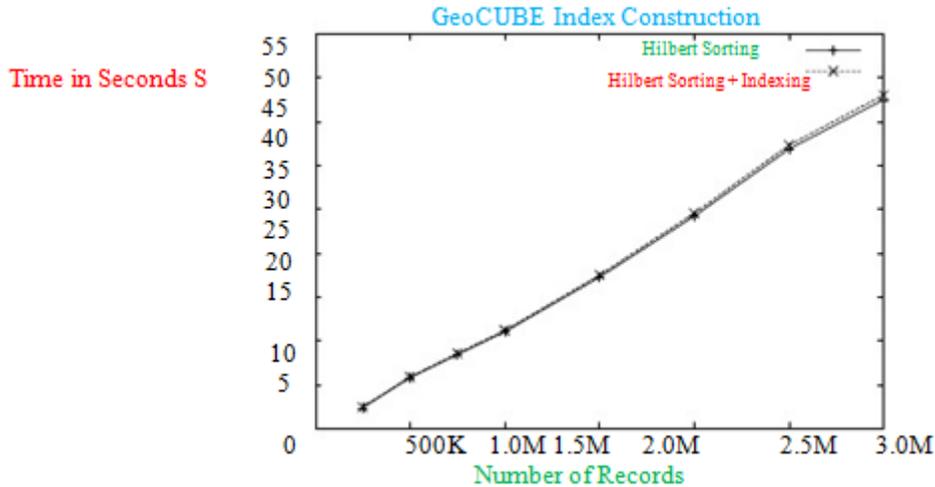

Figure 8: Time to construct the index including sorting of the dataset for the GeoCUBE index.

This implies the speed improvement of about 26% which can also be observed from the results for most other datasets tested, figure 8 shows the average range query time on the HYDRO implementation of the Indexing. The GCUBE query time increases only magically with an inverse in number of records and query results are reported in less than 0.2 seconds each from 3 m records. Note that some caution is acquired when interpreting their results. Since we are comparing two independent codes the bases it is wonder to determine how much of this improvement should be attributed to the improved I/O and cache efficiency of our algorithm and how much is due to better coding practices. We are exploring extensions our approach to support the indexing of the Spatial objects with extent as well as the use of compressed Hilbert indices. We also exploring how to support specific OLAP operations such as roll up and drill down on spatial dimensions.

## 5. CONCLUSION

In this research work, we presented a dynamic system for representation and indexing relational OLAP System with mixed categorical and continuous dimensions. The proposed approach is for representing and indexing relational OLAP with mixed categorical and continuous dimensions. The proposed approach is dynamic in nature and flexible enough with respect to dimension cardinality. Therefore, it allows from for the indexing of continuous space while building on top of established mechanisms for index construction, querying representation of mixed categorical /continuous data at storage level is the contribution of this research work. The outcome of the current research is the efficient SOLAP system that is capable of handling massive data. Empirical results imply the practical benefits of the proposed indexing approach. Hilbert curve space may also be applicable to multidimensional storage of Disks (MOLAP) as future direction of current research.

**AUTHORS**

Mr.M.Laxmaiah is research scholar in Jawaharlal Nehru Technological University, Kukatpally Hyderabad. He is currently working as Professor and Head of CSE Department in Tirumala Engineering College,Bogaram (V), Keesara (M), Hyderabad, AP, India. He has 15 Years of experience in Education and 4 Years of experience in Research field. He has 5 research Publications at International Journals. His areas of interest include Database Management Systems, Compiler Design and Data warehousing &Data Mining.

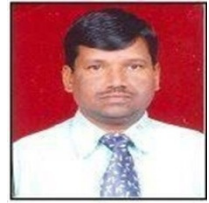

Dr.A.Govardhan did his BE in Computer Science and Engineering from Osmania University College of Engineering, Hyderabad in 1992.M.Tech from Jawaharlal Nehru University, Delhi in 1994 and PhD from Jawaharlal Nehru Technological University, Hyderabad in 2003.He is presently working as Professor in CSE, JNTUH, Kukatpally, Hyderabad. He has guided more than 120 M.Tech projects and number of MCA and B.Tech projects. He has 180 research publications at International/National Journals and Conferences. His areas of areas of interest include Databases, Data Warehousing &Mining, Information Retrieval, Computer Networks, Image processing and Object Oriented Technologies.

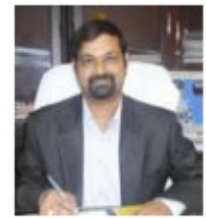